\documentclass[a4,12pt]{article}
\usepackage{float}
\usepackage{amsmath}
\def\ls{\lower0.5ex\hbox{$\buildrel >\over{\scriptstyle\sim}$}}
\def\rs{\lower0.5ex\hbox{$\buildrel <\over{\scriptstyle\sim}$}} 
\allowdisplaybreaks
\begin{document}
\pagestyle{empty} \setlength{\footskip}{2.0cm}
\setlength{\oddsidemargin}{0.5cm}
\setlength{\evensidemargin}{0.5cm}
\renewcommand{\thepage}{-- \arabic{page} --}
\def\mib#1{\mbox{\boldmath $#1$}}
\def\bra#1{\langle #1 |}  \def\ket#1{|#1\rangle}
\def\vev#1{\langle #1\rangle} \def\dps{\displaystyle}
\newcommand{\fcal}{{\cal F}}
\newcommand{\gcal}{{\cal G}}
\newcommand{\ocal}{{\cal O}}
\newcommand{\El}{E_\ell}
\renewcommand{\thefootnote}{$\sharp$\arabic{footnote}}
\newcommand{\W}{{\scriptstyle W}}
 \newcommand{\I}{{\scriptscriptstyle I}}
 \newcommand{\J}{{\scriptscriptstyle J}}
 \newcommand{\K}{{\scriptscriptstyle K}}
%
% ------------------------------------------------------------
 \def\thebibliography#1{\centerline{REFERENCES}
 \list{[\arabic{enumi}]}{\settowidth\labelwidth{[#1]}\leftmargin
 \labelwidth\advance\leftmargin\labelsep\usecounter{enumi}}
 \def\newblock{\hskip .11em plus .33em minus -.07em}\sloppy
 \clubpenalty4000\widowpenalty4000\sfcode`\.=1000\relax}\let
 \endthebibliography=\endlist
 \def\sec#1{\addtocounter{section}{1}\section*{\hspace*{-0.72cm}
 \normalsize\bf\arabic{section}.$\;$#1}\vspace*{-0.3cm}}
\def\secnon#1{\section*{\hspace*{-0.72cm}
 \normalsize\bf$\;$#1}\vspace*{-0.3cm}}
 \def\subsec#1{\addtocounter{subsection}{1}\subsection*{\hspace*{-0.4cm}
 \normalsize\bf\arabic{section}.\arabic{subsection}.$\;$#1}\vspace*{-0.3cm}}
% ------------------------------------------------------------
\vspace*{-1.7cm}
\begin{flushright}
$\vcenter{
%%% \phantom{  \hbox{{\footnotesize OUS and TOKUSHIMA Report}}  }
\hbox{{\footnotesize OUS and TOKUSHIMA Report}}
%%% \phantom{  {\hbox{{\footnotesize TOKUSHIMA Report}}}  }
%{ \hbox{(arXiv:1511.03437)}  }
}$
\end{flushright}

\vskip 1.6cm
\begin{center}
% \hspace*{-0.6cm}
  {\large \bf Refined analysis and updated constraints on general}
\vskip 0.18cm
  {\large\bf non-standard $\mib{tbW}$ couplings}

\end{center}

\vspace{0.5cm}
\begin{center}
\renewcommand{\thefootnote}{\alph{footnote})}
Zenr\=o HIOKI$^{\:1),\:}$\footnote{E-mail address:
\tt hioki@tokushima-u.ac.jp}\ 
Kazumasa OHKUMA$^{\:2),\:}$\footnote{E-mail address:
\tt ohkuma@ice.ous.ac.jp}\ and\
Akira Uejima$^{\:2),\:}$\footnote{E-mail address:
\tt uejima@ice.ous.ac.jp}\

\end{center}

\vspace*{0.4cm}
\centerline{\sl $1)$ Institute of Theoretical Physics,\ University of Tokushima}

\centerline{\sl Tokushima 770-8502, Japan}

\vskip 0.2cm
\centerline{\sl $2)$ Department of Information and Computer Engineering,}

\centerline{\sl Okayama University of Science}

\centerline{\sl Okayama 700-0005, Japan}

\vspace*{1.8cm}

\centerline{ABSTRACT}

\vspace*{0.2cm}
\baselineskip=21pt plus 0.1pt minus 0.1pt
We recently studied possible
non-standard $tbW$ couplings based on the effective-Lagrangian which consists of four kinds
of $SU(3)\times SU(2) \times U(1)$ invariant dimension-6 effective operators and gave an
experimentally allowed region for each non-standard coupling. We here re-perform that
analysis much more precisely based on the same experimental data but on a new computational
procedure using the Graphics-Processing-Unit (GPU) calculation system. Comparing these two
analyses with each other, the previous one is found to have given quite reliable results
despite of its limited computation capability. We then apply this new procedure to the
latest data and present updated results.

\vskip 1.5cm

\vfill
% PACS:  12.38.-t, 12.38.Bx, 12.38.Qk, 12.60.-i, 14.65.Ha, 14.70.Dj

% Keywords:
% Hadron colliders, Anomalous couplings, Top productions,\\ \ \ \ \ \ \  Lepton distributions
\setcounter{page}{0}
\newpage
\renewcommand{\thefootnote}{$\sharp$\arabic{footnote}}
%-------------------------------------------------------------
\pagestyle{plain} \setcounter{footnote}{0}

The top quark, the heaviest particle we have ever encountered up to now, is expected to
play an important role as a window opened for a possible new physics beyond the standard
model. The Large Hadron Collider (LHC) has been accumulating more and more data on this
quark and will soon enable its precision studies. In our recent article~\cite{Hioki:2015env},
we performed an analysis of possible non-standard top--bottom--$W$ ($tbW$) couplings as
model-independently as possible based on the effective-Lagrangian framework~
\cite{Buchmuller:1985jz}--\cite{Grzadkowski:2010es} by using available experimental data
of top-decay processes at the LHC.\footnote{We have given a detailed list of preceding
    works by other authors in~\cite{Hioki:2015env}. We would like to add \cite{Birman:2016jhg}
    to the list, which has appeared after our work.}

The effective-Lagrangian we used there consists of $SU(3)\times SU(2) \times U(1)$ invariant 
operators whose mass-dimension is six, and there are four kinds of operators that could
contribute to the $tbW$ couplings. There we have given allowed regions for those non-standard
couplings.
The precision level of the results was, however, not high enough due to its computational
limitation. In this note, we aim to re-analyze the same experimental data but on a new
computational procedure using the Graphics-Processing-Unit (GPU) calculation system.
We will thereby be able to check how reliable the last analysis was. We then apply this
procedure to the latest data and present more precise constraints on those couplings.

In our framework \cite{Hioki:2015env}, assuming that there exists some new physics characterized by
an energy scale ${\mit\Lambda}$ (e.g., the mass of a typical new particle) and all the non-standard
particles are not lighter than the LHC energy, the standard-model Lagrangian of $tbW$ interactions
describing phenomena around the electroweak scale is extended as
\begin{alignat}{1}\label{eq:efflag_decay}
  &{\cal L}_{tbW}  = -\frac{1}{\sqrt{2}}g 
  \Bigl[\,\bar{\psi}_b(x)\gamma^\mu(f_1^L P_L + f_1^R P_R)\psi_t(x)W^-_\mu(x) \Bigr.
  \nonumber\\
 &\phantom{========}
  +\bar{\psi}_b(x)\frac{\sigma^{\mu\nu}}{M_W}(f_2^L P_L + f_2^R P_R)
   \psi_t(x)\partial_\mu W^-_\nu(x) \,\Bigr],
\end{alignat}
where $g$ is the $SU(2)$ coupling constant, $P_{L/R}\equiv(1\mp\gamma_5)/2$, and $f_{1,2}^{L,R}$
stand for the corresponding coupling parameters. Among those parameters, we divide $f_1^L$ into
the SM term and the rest (i.e., the non-SM term) as 
\begin{equation}
f_1^L \equiv f_1^{\rm SM}+\delta\! f_1^L,
\label{eq:f1redef}
\end{equation}
where we assume $f_1^{\rm SM} (= V_{tb})=1$, and treat $\delta\! f_1^L$, $f_1^R$, and
$f_2^{L/R}$ as non-standard complex couplings which are all independent of each other.
%%%%%We handle the real and imaginary parts of all those non-standard couplings simultaneously.
%%%%%That is, we perform an eight-parameter analysis to find allowed regions for each coupling.

In order to give constraints on them, we use the following experimental information as our input
data:
\begin{quote}
$\bullet$ The total decay width of the top quark~\cite{Khachatryan:2014nda}
\end{quote}
\begin{equation}\label{eq:total_w}
{\mit\Gamma}^t = 1.36\pm 0.02({\rm stat.})^{+0.14}_{-0.11}({\rm syst.}) ~~{\rm GeV}.\footnote{
    In fact, it is not easy to handle an asymmetric error like this in the error propagation.
    We therefore use ${\mit\Gamma}^t = 1.36\pm 0.02({\rm stat.})\pm 0.14({\rm syst.})~{\rm GeV}$,
    the one symmetrized by adopting the larger (i.e., $+0.14$) in this systematic error.}
\end{equation}
\begin{quote}
$\bullet$ The partial decay widths derived from experimental data of $W$-boson helicity
fractions~\cite{Khachatryan:2014vma} with the above ${\mit\Gamma}^t$
\end{quote}

\vspace*{-0.7cm}
\begin{equation}\label{eq:gamma_eff}
\begin{split}
 &{\mit\Gamma}_L^{t*}=0.405\pm 0.072~{\rm GeV},\\
 &{\mit\Gamma}_0^{t*}=0.979\pm 0.125 ~{\rm GeV},\\
 &{\mit\Gamma}_R^{t*}=-0.024\pm0.030~{\rm GeV}.
\end{split}
\end{equation}

\vskip 0.2cm \noindent
Varying all the parameters at the same time, we look for the area in the parameter space in which
we find solutions to satisfy the above input and outside of which any parameter values there do not
reproduce the data. We then represent the resultant allowed region for each parameter by giving
its maximum and minimum values. Throughout the computations, we do not neglect
any contributions, i.e., we keep not only the SM term plus those linear in the non-standard couplings
but also those quadratic in them.

In the previous work~\cite{Hioki:2015env}, the analysis was carried out by varying each parameter
in steps of 0.05 using a workstation $[\ {\rm 67.2GFLOPS}\ ]$. Here we re-analyze the same data
in order to see if we could give more precise constraints on each parameter using a GPU calculator
$[\ {\rm 4.29TFLOPS}\ ]$.
We take as $m_t=172.5~{\rm GeV}$, $m_b=4.8~{\rm GeV}$ and $M_W=80.4~{\rm  GeV}$
for the masses of the involved particles as in~\cite{Hioki:2015env}.

The results corresponding to the previous ones are shown in Tables \ref{tab:8para}, \ref{tab:7para}
and \ref{tab:6para}: the allowed regions between the maximum and minimum in those tables have been
obtained respectively from the eight-parameter analysis (i.e. all the parameters are treated as
free ones) in steps of 0.02, the seven-parameter one (i.e. Re$(\delta\! f_1^L)=0$, the others are
treated as free parameters) and the six-parameter one (i.e. Re$(\delta\! f_1^L)
={\rm Im} (\delta\! f_1^L)=0$, the others are treated as free parameters) both
in 0.01 steps.\footnote{It would take more than 12 years to get a meaningful result in
    an eight-parameter analysis in steps of 0.01, even if the GPU calculator were used.
    Therefore, we have adopted 0.02 steps for the eight-parameter analysis.}\ \

\vspace{0.6cm}

\begin{table}[H]
\centering
\caption{Allowed maximum and minimum values of the non-standard-top-decay couplings in the case
that all the couplings are dealt with as free parameters. Those in the parentheses show the previous
results.}
\label{tab:8para}
\vspace*{0.4cm}
\begin{tabular}{c|cc|cc|cc|cc}
& \multicolumn{2}{c|}{$\delta\! f_1^L$}& \multicolumn{2}{c|}{$f_1^R$}
& \multicolumn{2}{c|}{$f_2^L$}& \multicolumn{2}{c}{$f_2^R$}
\\ \cline{2-9} 
& Re($\delta\! f_1^L$)
&\hspace*{-0.4cm} Im($\delta\! f_1^L$) & Re($f_1^R$)
&\hspace*{-0.4cm} Im($f_1^R$) & Re($f_2^L$)
&\hspace*{-0.4cm} Im($f_2^L$) & Re($f_2^R$)
&\hspace*{-0.4cm} Im($f_2^R$)
\\ \cline{1-9}
%----------
Min. & $-2.58$    &\hspace*{-0.4cm} $-1.58$           & $-1.36$           
&\hspace*{-0.4cm} $-1.36$           & $-0.68$           
&\hspace*{-0.4cm} $-0.68$           & $-1.20$            
&\hspace*{-0.4cm} $-1.20$           \\ %\hline
%----------
     & ${\scriptstyle(-2.55)}$  &\hspace*{-0.4cm} ${\scriptstyle(-1.55)}$  & ${\scriptstyle(-1.30)}$           
&\hspace*{-0.4cm} ${\scriptstyle(-1.30)}$  & ${\scriptstyle(-0.65)}$           
&\hspace*{-0.4cm} ${\scriptstyle(-0.65)}$  & ${\scriptstyle(-1.20)}$            
&\hspace*{-0.4cm} ${\scriptstyle(-1.20)}$         \\ %\hline
%---------- ----------
Max. & $\phantom{-}0.58$ &\hspace*{-0.4cm} $\phantom{-}1.58$ & $\phantom{-}1.36$ 
&\hspace*{-0.4cm} $\phantom{-}1.36$ & $\phantom{-}0.68$ 
&\hspace*{-0.4cm} $\phantom{-}0.68$  & $\phantom{-}1.20$  
&\hspace*{-0.4cm} $\phantom{-}1.20$ \\
%----------
     & $\phantom{-}{\scriptstyle(0.55)}$ &\hspace*{-0.4cm} $\phantom{-}{\scriptstyle(1.55)}$ 
     & $\phantom{-}{\scriptstyle(1.30)}$ 
&\hspace*{-0.4cm} $\phantom{-}{\scriptstyle(1.30)}$ & $\phantom{-}{\scriptstyle(0.65)}$ 
&\hspace*{-0.4cm} $\phantom{-}{\scriptstyle(0.65)}$ & $\phantom{-}{\scriptstyle(1.20)}$  
&\hspace*{-0.4cm} $\phantom{-}{\scriptstyle(1.20)}$
%----------
\end{tabular}
\end{table}

\vspace{0.3cm}

\begin{table}[H]
\centering
\caption{Allowed maximum and minimum values of non-standard-top-decay couplings in the case
that all the couplings are dealt with as free parameters except for Re$(\delta\! f_1^L)$ being
set to be zero. Those in the parentheses show the previous results.}
\label{tab:7para}
\vspace*{0.4cm}
\begin{tabular}{c|c|cc|cc|cc}
& {$\delta\! f_1^L$}& \multicolumn{2}{c|}{$f_1^R$}& \multicolumn{2}{c|}{$f_2^L$}
& \multicolumn{2}{c}{$f_2^R$}             \\ \cline{2-8}
&  Im($\delta\! f_1^L$) & Re($f_1^R$) & Im($f_1^R$) & Re($f_2^L$) &Im($f_2^L$) 
& Re($f_2^R$) & Im($f_2^R$) \\  \cline{1-8}
Min. 	&  $-1.23$  & $-1.14$ 
	& $-1.12$ & $-0.55$ 
        & $-0.57$ & $-0.96$ 
        & $-1.00$ \\ %\hline
 %----------
&\hspace*{-0.0cm} ${\scriptstyle(-1.20)}$  
&\hspace*{-0.0cm} ${\scriptstyle(-1.10)}$  &\hspace*{-0.0cm} ${\scriptstyle(-1.10)}$
&\hspace*{-0.0cm} ${\scriptstyle(-0.50)}$  &\hspace*{-0.0cm} ${\scriptstyle(-0.55)}$
&\hspace*{-0.0cm} ${\scriptstyle(-0.95)}$  &\hspace*{-0.0cm} ${\scriptstyle(-1.00)}$
\\ %\hline
%---------- ----------       
Max. 
& $\phantom{-}1.23$  
& $\phantom{-}1.10$ & $\phantom{-}1.12$
& $\phantom{-}0.59$  & $\phantom{-}0.57$
& $\phantom{-}0.00$ & $\phantom{-}1.00$\\
%----------
&\hspace*{-0.1cm} $\phantom{-}{\scriptstyle(1.20)}$  
&\hspace*{-0.1cm} $\phantom{-}{\scriptstyle(1.05)}$  &\phantom{-}\hspace*{-0.0cm} ${\scriptstyle(1.10)}$
&\hspace*{-0.1cm} $\phantom{-}{\scriptstyle(0.55)}$  &\phantom{-}\hspace*{-0.0cm} ${\scriptstyle(0.55)}$
&\hspace*{-0.1cm} $\phantom{-}{\scriptstyle(0.00)}$  &\phantom{-}\hspace*{-0.0cm} ${\scriptstyle(1.00)}$
\\ %\hline
\end{tabular}
\end{table}

\vspace{0.3cm}

\begin{table}[H]
\centering
\caption{Allowed maximum and minimum values of non-standard-top-decay couplings in the case
that all the couplings are dealt with as free parameters except for Re$(\delta\! f_1^L)$ and
Im$(\delta\! f_1^L)$ both being set to be zero. Those in the parentheses show the previous results.}
\label{tab:6para}
\vspace*{0.4cm}
\begin{tabular}{c|cc|cc|cc}
& \multicolumn{2}{c|}{$f_1^R$}& \multicolumn{2}{c|}{$f_2^L$}
& \multicolumn{2}{c}{$f_2^R$} \\ \cline{2-7} 
&  Re($f_1^R$) & Im($f_1^R$) & Re($f_2^L$) &Im($f_2^L$) & Re($f_2^R$) & Im($f_2^R$) \\ \cline{1-7}
Min. 	&  $-1.14 $ & $-1.12$ 
 	& $-0.55$ & $-0.57$ 
        & $-0.96$ & $-0.49$  \\ %\hline
 %----------
&\hspace*{-0.1cm} ${\scriptstyle(-1.10)}$ &\hspace*{-0.1cm} ${\scriptstyle(-1.10)}$  
&\hspace*{-0.1cm} ${\scriptstyle(-0.50)}$ &\hspace*{-0.1cm} ${\scriptstyle(-0.55)}$
&\hspace*{-0.1cm} ${\scriptstyle(-0.95)}$ &\hspace*{-0.1cm} ${\scriptstyle(-0.45)}$  \\ %\hline        
Max. 	&  $\phantom{-}1.10$   & $\phantom{-}1.12$ 
	&  $\phantom{-}0.59$   & $\phantom{-}0.57$
     	&  $\phantom{-}0.00$   & $\phantom{-}0.49$\\
         %----------
&\hspace*{-0.1cm} $\phantom{-}{\scriptstyle(1.05)}$ &\phantom{-}\hspace*{-0.0cm} ${\scriptstyle(1.10)}$  
&\hspace*{-0.1cm} $\phantom{-}{\scriptstyle(0.55)}$ &\phantom{-}\hspace*{-0.0cm} ${\scriptstyle(0.55)}$
&\hspace*{-0.1cm} $\phantom{-}{\scriptstyle(0.00)}$ &\phantom{-}\hspace*{-0.0cm} ${\scriptstyle(0.45)}$ \\
\end{tabular}
\end{table}

%%%%%\vspace{0.5cm}

From a general point of view, the maximum/minimum of the allowed region is expected to
increase/decrease by up to 0.05 (0.04) if we change the step size from 0.05 to 0.01 (0.02).
The actual changes of the boundaries are however smaller than this naive expectation except
for that of $f_1^R$ in Table 1. The fact that most of them have not changed so much means
that our previous analysis has already given quite reliable results despite of its rather large
step size. In addition, the exceptional behavior of $f_1^R$ tells us that several parameters
could interact with each other in analyses like the present one and consequently some parameters
get larger allowed regions than we imagine. Let us note that this would never happen in a
``multiple-parameter analysis'' in which only one parameter is varied at once.

This way we have confirmed that our previous analysis is well reliable, but it does not mean that we have
obtained nothing new in the present analysis. In order to show how the precision has been
raised here, we give in Table \ref{tab:rate} the increase rate of each allowed region in percentage.
We see that this re-analysis has been worth performing especially for $f_1^R$ and $f_2^L$.
It is also remarkable that the increase rates for Re($f_2^R$) are quite small in contrast to the
other parameters. The reason will be that the change of Re($f_2^R$) is not cancelled out by
contributions from the other parameters, because that from Re($f_2^R$) on the $tbW$ couplings is
the largest one (except for the one from the standard model) since only this term can interfere
with the standard-model term when the $b$-quark mass is neglected, i.e., all the other interference terms
are proportional to $m_b$.

\vspace{0.5cm}

\begin{table}[H]
\caption{The increase rates of the allowed regions compared with the previous results.}
\label{tab:rate}
\vspace*{0.4cm}
\begin{tabular}{c|cc|cc|cc|cc}
& \multicolumn{2}{c|}{$\delta\! f_1^L$}& \multicolumn{2}{c|}{$f_1^R$}
& \multicolumn{2}{c|}{$f_2^L$}& \multicolumn{2}{c}{$f_2^R$}
\\ \cline{2-9} 
&\hspace*{-0.25cm} Re($\delta\! f_1^L$)
&\hspace*{-0.45cm} Im($\delta\! f_1^L$)
&\hspace*{-0.25cm} Re($f_1^R$)
&\hspace*{-0.45cm} Im($f_1^R$)
&\hspace*{-0.25cm} Re($f_2^L$)
&\hspace*{-0.45cm} Im($f_2^L$)
&\hspace*{-0.25cm} Re($f_2^R$)
&\hspace*{-0.45cm} Im($f_2^R$)
\\ \cline{1-9}
8 param.%anal.
    & 1.9\%   & 1.9\%   & 4.6\%    &4.6\%    & 4.6\%    & 4.6\%    & 0.0\% & 0.0\% \\ %\hline
7 param.%anal.
    &   ---   & 2.5\%   & 4.2\%    &1.8\%    & 8.6\%    & 3.6\%    & 1.1\% & 0.0\% \\
6 param.%anal.
    &   ---   &   ---   & 4.2\%    &1.8\%    & 8.6\%    & 3.6\%    & 1.1\% & 8.9\%
\end{tabular}
\end{table}

\vspace{0.5cm}

Now we know that our strategy and procedure are trustable and therefore we are ready to refine
the analysis based on the following latest data on the partial decay widths:

\begin{quote}\vspace*{-0.2cm}
$\bullet$ The partial decay widths derived from experimental data of $W$-boson helicity
fractions~\cite{Khachatryan:2016fky} with ${\mit\Gamma}^t$ in eq.(\ref{eq:total_w})
\end{quote}
\begin{equation}\label{eq:gamma_eff-new}
\begin{split}
 &{\mit\Gamma}_L^{t*}=\phantom{-}0.439\pm 0.051~{\rm GeV},\\
 &{\mit\Gamma}_0^{t*}=\phantom{-}0.926\pm 0.103 ~{\rm GeV},\\
 &{\mit\Gamma}_R^{t*}=-0.005\pm0.020~{\rm GeV}.
\end{split}
\end{equation}

\vskip 0.4cm
We present the results of the eight-, seven- and six-parameter analyses
in Tables \ref{tab:8para-new}, \ref{tab:7para-new}, and \ref{tab:6para-new} respectively,
which should be compared with those in Tables \ref{tab:8para}, \ref{tab:7para} and
\ref{tab:6para}. We see that there are meaningful improvements of 0.03--0.04 in a couple of
results for $\delta\! f_1^L$ and $f_1^R$. Although some other boundaries have also changed,
their sizes are of 0.02, which might come from the difference between the central values of
(\ref{eq:gamma_eff}) and (\ref{eq:gamma_eff-new}) and therefore it is not easy yet to draw
definite conclusion on them. In any case, these results are all consistent with the
standard-model predictions, but it is also noteworthy that there still exists enough space
for a new physics beyond the standard model.

\vspace{0.5cm}

%%%%% 8-parameters
\begin{table}[H]
\centering
\caption{Updated constraints on the non-standard-top-decay couplings in the case
that all the couplings are dealt with as free parameters.}
\label{tab:8para-new}
\vspace*{0.4cm}
\begin{tabular}{c|cc|cc|cc|cc}
& \multicolumn{2}{c|}{$\delta\! f_1^L$}& \multicolumn{2}{c|}{$f_1^R$}
& \multicolumn{2}{c|}{$f_2^L$}& \multicolumn{2}{c}{$f_2^R$}
\\ \cline{2-9} 
& Re($\delta\! f_1^L$)
&\hspace*{-0.4cm} Im($\delta\! f_1^L$) & Re($f_1^R$)
&\hspace*{-0.4cm} Im($f_1^R$) & Re($f_2^L$)
&\hspace*{-0.4cm} Im($f_2^L$) & Re($f_2^R$)
&\hspace*{-0.4cm} Im($f_2^R$)
\\ \cline{1-9}
Min. & $-2.56$           &\hspace*{-0.4cm} $-1.56$           & $-1.32$           
&\hspace*{-0.4cm} $-1.32$           & $-0.70$           
&\hspace*{-0.4cm} $-0.70$           & $-1.20$            
&\hspace*{-0.4cm} $-1.20$           \\ %\hline
Max. & $\phantom{-}0.56$ &\hspace*{-0.4cm} $\phantom{-}1.56$ & $\phantom{-}1.32$ 
&\hspace*{-0.4cm} $\phantom{-}1.32$ & $\phantom{-}0.70$ 
&\hspace*{-0.4cm} $\phantom{-}0.70$  & $\phantom{-}1.20$  
&\hspace*{-0.4cm} $\phantom{-}1.20$
\end{tabular}
%}
\end{table}

\vspace{0.3cm}

%%%%% 7-parameters
\begin{table}[H]
\centering
\caption{Updated constraints on the non-standard-top-decay couplings in the case
that all the couplings are dealt with as free parameters except for Re$(\delta\! f_1^L)$ being
set to be zero.}
\label{tab:7para-new}
\vspace*{0.4cm}
\begin{tabular}{c|c|cc|cc|cc}
& {$\delta\! f_1^L$}& \multicolumn{2}{c|}{$f_1^R$}& \multicolumn{2}{c|}{$f_2^L$}
& \multicolumn{2}{c}{$f_2^R$}             \\ \cline{2-8}
&  Im($\delta\! f_1^L$) & Re($f_1^R$) & Im($f_1^R$) & Re($f_2^L$) &Im($f_2^L$) 
& Re($f_2^R$) & Im($f_2^R$) \\  \cline{1-8}
Min. &  $-1.20$  & $-1.12$ & $-1.10$ & $-0.57$ & $-0.59$ & $-0.96$ & $-1.00$ \\ %\hline
Max. &  $\phantom{-}1.20$  & $\phantom{-}1.07$ & $\phantom{-}1.10$ & $\phantom{-}0.61$ 
     & $\phantom{-}0.59$   & $\phantom{-}0.00$ & $\phantom{-}1.00$
\end{tabular}
\end{table}

\newpage % \vspace{0.3cm}

%%%%% 6-parameters
\begin{table}[H]
\centering
\caption{Updated constraints on the non-standard-top-decay couplings in the case
that all the couplings are dealt with as free parameters except for Re$(\delta\! f_1^L)$ and
Im$(\delta\! f_1^L)$ both being set to be zero.}
\label{tab:6para-new}
\vspace*{0.4cm}
\begin{tabular}{c|cc|cc|cc}
& \multicolumn{2}{c|}{$f_1^R$}& \multicolumn{2}{c|}{$f_2^L$}
& \multicolumn{2}{c}{$f_2^R$} \\ \cline{2-7} 
&  Re($f_1^R$) & Im($f_1^R$) & Re($f_2^L$) &Im($f_2^L$) & Re($f_2^R$) & Im($f_2^R$) \\ \cline{1-7}
Min. &  $-1.12 $ & $-1.10$ & $-0.57$ & $-0.59$ & $-0.96$ & $-0.49$  \\ %\hline
Max. &  $\phantom{-}1.07$  & $\phantom{-}1.10$ & $\phantom{-}0.61$ & $\phantom{-}0.59$
     & $\phantom{-}0.00$   & $ \phantom{-}0.49$
\end{tabular}
\end{table}

\vspace{0.4cm}
In conclusion, we have performed a re-analysis of the same experimental data as the previous one
but on a new computational procedure, and presented more precise allowed regions of the non-standard $tbW$ 
couplings treated as complex numbers. There we find that the allowed regions have become slightly
larger than those of the previous analysis. However, the overall tendencies of these two analyses
seem to be consistent with each other. We therefore would like to stress that we have succeeded to
raise the precision level through the re-analysis here and also that the previous analysis was better
than we naively imagine from its limited-precision calculations. Having confirmed this way that our strategy
and procedure are trustable, we then have made a new analysis based on the latest experimental data and
given updated constraints on those couplings. There the constraints on $\delta\! f_1^L$ and $f_1^R$ have
been further improved, and some other boundaries have also shown certain small changes.

%%%%%%%%%%%%%%%%%%%%%%%%%%%%%%%%%%%%%%%%%%%%%%%%%%%%%%%%%%%%%%%%%%%%%%%%%%
%
\secnon{Acknowledgments}
%
%%%%%%%%%%%%%%%%%%%%%%%%%%%%%%%%%%%%%%%%%%%%%%%%%%%%%%%%%%%%%%%%%%%%%%%%%%%%%%
Part of the algebraic and numerical calculations in the early stage 
were carried out on the computer system at Yukawa Institute for
Theoretical Physics (YITP), Kyoto University.

\baselineskip=20pt plus 0.1pt minus 0.1pt

\vspace*{0.8cm}

% RRRRRRRRRRRRRRRRRRRRRRRRRRRRRRRRRRR

\end{document}